\documentclass[aps,prd,email,showpacs,twocolumn,preprintnumbers,showkeys,superscriptaddress,amsmath,amssymb,nofootinbib]{revtex4-2}

\usepackage{bbold}
\usepackage{color}
\usepackage{latexsym}
\usepackage{amsmath}
\usepackage{amssymb}
\usepackage[utf8]{inputenc}
\usepackage{amsmath}
\usepackage{amsfonts}
\usepackage{bm}
\usepackage{bbold}
\usepackage{braket}
\usepackage{amssymb}
\usepackage{epstopdf}
\usepackage{epsfig}
\usepackage{eufrak}
\usepackage{euscript}
\usepackage{pstricks}
\usepackage{graphics}
\usepackage{graphicx}
\usepackage{float}	
\usepackage{picture}
\usepackage{appendix}
\usepackage{enumitem}
\usepackage{dsfont}

\newcommand{\be}{\begin{equation}}
\newcommand{\ee}{\end{equation}}
\newcommand{\ba}{\begin{eqnarray}}
\newcommand{\ea}{\end{eqnarray}}

\usepackage{hyperref}
\hypersetup{
    colorlinks=true,
    linkcolor=blue,
    filecolor=teal,
    urlcolor=blue,
    citecolor=teal
}

\DeclareFontEncoding{LS1}{}{}
\DeclareFontSubstitution{LS1}{stix}{m}{n}
\DeclareSymbolFont{stixletters}{LS1}{stix}{m}{it}
\DeclareMathAccent{\cev}{\mathord}{stixletters}{"91}
\DeclareMathAccent{\vec}{\mathord}{stixletters}{"92}
\DeclareMathAccent{\vecev}{\mathord}{stixletters}{"95}
\begin{document}
\title{\Large{Loop Quantum Gravity effects on electromagnetic properties \\ of charged leptons}}
\author{ J. P. S. Melo } \email{jpaulosm@cbpf.br}
\affiliation{Centro Brasileiro de Pesquisas F\'isicas, Rua Dr. Xavier Sigaud 150, Urca, CEP 22290-180, Rio de Janeiro, RJ, Brazil}
\author{M. J. Neves}\email{mariojr@ufrrj.br}
\affiliation{Departamento de F\'isica, Universidade Federal Rural do Rio de Janeiro, BR 465-07, CEP 23890-971, Serop\'edica, RJ, Brazil}
\author{J. M. A. Paix\~ao} \email{jeff@cbpf.br}
\affiliation{Centro Brasileiro de Pesquisas F\'isicas,  Rua Dr. Xavier Sigaud 150, Urca, CEP 22290-180, Rio de Janeiro, RJ, Brazil}
\author{J. A.  Helay\"el-Neto}\email{helayel@cbpf.br}
\affiliation{Centro Brasileiro de Pesquisas F\'isicas, Rua Dr. Xavier Sigaud 150, Urca, CEP 22290-180, Rio de Janeiro, RJ, Brazil}

\begin{abstract}
The efforts in this contribution consist in reassessing a modified Dirac equation that incorporates a $\gamma^0 \gamma_5$-Lorentz-symmetry violating (LSV) term induced as a Loop Quantum Gravity (LQG) effect. Originally, this equation has been applied and considered as a good scenario for describing a number of investigations on the flight time of cosmic neutrinos, which suggests that the speed, in vacuum, in connection with the geometry that describes a granular space-time, takes the energy-dependent form, {\it e.g.}, $v(E) =1 \pm  E/E_{\textrm{\tiny LSV}}$, with $E_{\textrm{\tiny LSV}} \approx 6,5 \times 10^{17}$ GeV for neutrinos. Once LQG provides a viable way to understand this picture consistently, we pursue an analysis of this effective Dirac equation to inspect some of its properties. These include: the derivation of the modified fermionic propagator, attainment of the Gordon decomposition of the vector current with minimal electromagnetic coupling to obtain information on the form factors, examination of the non-relativistic limit of the equation, evaluation of the spin- and velocity-dependent corrections to the Coulomb potential due to LQG effects, and the modified Hamiltonian in the low-relativistic regime. The study of the form factors may open up paths to set up bounds on the LQG parameters from the precision measurements of electromagnetic attributes of the charged leptons, such as their respective electric and magnetic dipole moments.
\end{abstract}
\maketitle
\newpage
\pagestyle{myheadings}
\section{Introduction} \label{sec1}
The theoretical framework of Loop Quantum Gravity (LQG) is one of the most promising ways to inspect the challenges  of quantum gravity, which seeks to understand the fundamental nature of the spacetime and gravity at the quantum level \cite{Rovelli98, Ashtekar04}. In LQG, the spacetime is considered to be discrete rather than continuous, which is a departure from the smooth, continuous spacetime described by general relativity. The theory represents the spacetime as a network or quantum foam of interconnected loops. These loops are thought to be the building blocks of space and time at the smallest scales. The key idea in LQG is that quantities, such as area and volume, are quantized, meaning they come in discrete, indivisible units. The theory uses mathematical structures called spin networks and spin foams to describe the quantum properties of the spacetime \cite{Ashtekar86, Jacobson87, Rovelli88, Ashtekar92, Rovelli94a, Rovelli94b, Rovelli95, Rovelli96, Thiemann96, Baez96, Fort97, Perez13, Han19}. LQG is still a research area in progress and it has not yet been  experimentally confirmed once its principal predictions and phenomenological aspects become relevant at the order of the Planck scale ($ \ell_{P} \sim \!10^{19}$ GeV) \cite{Gambini99, Alfaro02, Alfaro02b}. However, indirect ways to probe the soft effects of quantum gravity at larger scales or through cosmological observations, {\it e.g}, some experiments and observations related to cosmology \cite{Graef20, Barboza22}, high-energy astrophysics \cite{Amelino-Camelia98, Vagnozzi22, Afrin22, Yan23, Chen24}, or precision measurements \cite{Zhu20, Anderson23} might provide insights into LQG phenomenology.
The investigation on Lorentz-symmetry violation (LSV) is another significant approach to uncover evidence in the pursuit of quantum gravity, once it is also expected to happen at the Planck scale \cite{Amelino-Camelia98, Amelino97, Ellis99,  Gambini99,  Coleman99}, where strong space-time fluctuations are expected to occur. In this context, discussions related to tests based on Modified Dispersion Relations (MDRs), using high-energy photons available from astrophysical measurements, can contribute to analyses of possible deviations compared to standard dispersion relations, indicating the presence of LSV or some effects of quantum gravity, such as LQG \cite{Amelino-Camelia98, Amelino97, Ellis99}. Additionally, in this context, an important connection between the LQG and LSV frameworks is presented in \cite{Gambini99}, where the modified Maxwell equations point to a vacuum birefringence. Two other important works regarding the MDRs, characterizing dispersion relations that could indicate the existence of the LSV, can be found in Refs. \cite{Amelino00, Alfaro02}.
In the context of MDRs, as argued in \cite{Li23} at the leading order of LSV, the modified speed of a massless particle with energy $E$ can be expressed as $v(E) \approx (1 - \xi E)$, where $\xi$ refers to the LSV scale approximately of the order of a fraction of the Planck scale. Consequently, the velocity difference between two massless particles with different energies becomes apparent whenever their energies are sufficiently high, and the discrepancies in travel time accumulate over a considerable propagation distance. As a result, cosmic photons and neutrinos, particularly those originating from gamma-ray bursts (GRBs), and active galactic nuclei (AGNs) exhibiting simultaneous high energy, provide a unique window for exploring LSV during their journeys through the Universe \cite{Amelino-Camelia98}.  Regarding the LSV exploration linked to GRBs, a comprehensive summary is provided in \cite{Li23} stating  that: for photons, without helicity dependence, we have $v(E) = (1 - E/E^\gamma_{\textrm{\tiny LSV}})$, with $E^\gamma_{\textrm{\tiny LSV}} \approx 3.6 \times 10^{17}$ GeV, indicating that photons are subluminal and high energy photons propagate slower than low energy ones \cite{Shao10, Zhang15, Xu16a, Xu16b, Amelino-Camelia17, Liu18, Li20, Zhu21, Chen21, Xu18, Amelino-Camelia21}; for neutrinos and anti-neutrinos, their speeds are $v(E) \approx (1 \mp E/E^{\nu}_{\textrm{\tiny LSV}})$, with $E^{\nu}_{\textrm{\tiny LSV}} \approx  6.5 \times 10^{17}$ GeV, indicating that neutrinos and anti-neutrinos, or Majorana neutrinos with opposite helicities, could be subluminal and superluminal with the sign of the velocity variations depending on the helicities \cite{Xu18, Amelino-Camelia21, Amelino-Camelia15, Amelino-Camelia16, Huang18, Huang19, Huang22}.
In \cite{Li23} the authors find out that LQG is as a good theoretical candidate for explaining the phenomenological suggestion of the speed variations of cosmic photons and neutrinos picture consistently, and suggest that other observable signals could also testify the predictions of LQG in the future. In order to show that, they introduce LQG description that consists in assuming specifics conditions and properties for the so-called weave states, which are semiclassical states with a characteristic length $\mathcal{L}$ that describes the discreteness of the spacetime represented by this state. When the de Broglie wavelength $(\lambda)$ of a particle satisfies the condition $\ell_{\textrm{\tiny P}} \ll \lambda$, in which $\ell_{\textrm{\tiny P}}=1.61\times 10^{-35}$ m is the Planck length, and the characteristic length $\mathcal{L}$ satisfies the condition $\ell_{\textrm{\tiny P}} \ll \mathcal{L} \le \lambda$, one can compute the weave state expectation of the corresponding Hamiltonian of the particle and obtain an effective Hamiltonian from which we can read off the modified dispersion relation and the speed of such particle \cite{Alfaro02, Alfaro02b}. Their results can be directly related to observable phenomena and conversely the phenomenological analyses may be used to examine the description of LQG based on weave states, leading towards better understanding of the LQG framework and further progress of LQG theories.
Our efforts in this contribution consist in reassessing a modified Dirac equation that incorporates a $\gamma^0 \, \gamma_5$-LSV term induced as a LQG effect by \cite{Li23}. The paper is structured as follows: In Section \ref{sec2}, we present the modified Dirac equation discussed in \cite{Li23}. We provide the action, and the modified fermion propagator associated with this equation. The dispersion relation is reevaluated, and additionally, we introduce the correspondent group velocity.  In Section \ref{sec3}, we derive the positive and negative energy solutions for this modified Dirac equation.  The Section \ref{sec4} focuses on obtaining the conserved current associated with the $U(1)$ gauge symmetry of the model. Through minimal coupling with the electromagnetic (EM) field, we demonstrate that this current aligns with the EM current within the framework of modified Maxwell's equations. In Section \ref{sec5}, we obtain the Gordon decomposition of the conserved current exploring its implications. Section \ref{sec6} is devoted to obtaining spin- and velocity-dependent corrections to the Coulomb potential due to the modified current structure of the model. The Section \ref{sec7} involves deriving the non-relativistic limit of the modified Dirac equation with minimal coupling.  In Section \ref{sec8}, we delve into the question of renormalizability, concluding that the model is super-renormalizable. Finally, Section \ref{sec9} encompasses our concluding considerations.
We adopt the natural system of units, in which $c = \hbar =1$, and we use the Minkowski signature $(+, -, -, -)$.
\section{The modified Dirac equation from LQG} \label{sec2}
The modified Dirac equation emerging from the LQG effects in the context of \cite{Li23} is
\begin{align}
\left(i\gamma^\mu\partial_\mu -\dfrac{\hat{C}}{2\mathcal{L}} \, \gamma^0 \gamma_5 - m \right)\psi = 0 \; ,
\label{Dirac1}
\end{align}
where $\mathcal{L}$ is a characteristic length constraint by the condition ${\ell}_{\textrm{\tiny P}} \ll \mathcal{L} \leqslant \lambda$,
in which $\ell_{\textrm{\tiny P}}=1.61\times 10^{-35}$ m is the Planck length, and $\lambda$ is the de Broglie wavelength of a particle. The
$\hat{C}$-operator is defined by
\begin{align} \label{Coperator}
\hat{C} = \dfrac{ \kappa_7}{2} \left(\dfrac{\ell_p}{\mathcal{L}} \right)^{\!\! \Upsilon_{\!f}} \!\! \ell_p^2 \, \bm \nabla^2 = \xi \, \bm \nabla^2 \; ,
\end{align}
in which $\Upsilon_f$ is a phenomenological parameter from LQG, $\kappa_7$ is a magnitude parameter, and $\xi$ is the combination of these parameters that has length dimension. Thus, it is convenient to define the quantity $\vartheta:=\xi/(2\mathcal{L})$, with length dimension, and consequently, all the known results of the fermion theory are recovered in the limit $\vartheta \rightarrow 0$. The $\gamma^{\mu}$-matrices are the usual $4\times 4$ Dirac's matrices that satisfy the relations $\gamma^{\mu}\gamma^{\nu}=\eta^{\mu\nu}\,\mathds{1}-i\,2\Sigma^{\mu\nu}$, and $\left\{ \gamma^{\mu},\gamma_5\right\}=0$. The correspondent action from Eq. (\ref{Dirac1}) is
\begin{align}\label{ActionDirac1}
{\cal S}(\bar{\psi},\psi) = \! \int  \! d^4x \,  \bar{\psi} \left(i\gamma^\mu\partial_\mu -\vartheta \, \gamma^0 \gamma_5\,\bm \nabla^2 - m\right)\psi \; ,
\end{align}
that, upon an integration by parts, we obtain
\begin{align}
{\cal S}(\bar{\psi},\psi) \!  = \! \int \!  d^4x \! \left[ \bar{\psi} \left(i\gamma^\mu\partial_\mu  \! - \! m\right)\psi
\! + \! \vartheta \, (\bm{\nabla} \bar{\psi}) \! \cdot \! \gamma^0 \gamma_5 (\bm{\nabla} \psi)  \right]  ,
\end{align}
where $\bar{\psi}:=\psi^{\dagger}\gamma^{0}$ is the adjoint field that satisfies the equation
\begin{eqnarray}\label{eqAdjoint}
i\,\partial_{\mu}\bar{\psi}\,\gamma^{\mu}+\vartheta\,(\bm \nabla^2\bar{\psi})\,\gamma^0 \gamma_5+m\,\bar{\psi}=0 \; .
\end{eqnarray}
Taking a plane wave solution $\psi(x)=u(p)\,e^{-ip\cdot x}$, the modified Dirac equation in momentum space reads
\begin{align}
\left( \gamma^\mu p_{\mu} +\vartheta\, \bm{p}^2 \gamma^0 \gamma_5 - m \right)u(p)= 0 \; ,
\label{eqD1}
\end{align}
where $u(p)$ is an amplitude in the momentum space.
The term $\vartheta \, \bm{p}^2$ appearing alongside the $\gamma^0\gamma_5$-matrix in Eq. \eqref{eqD1}, is a CPT-violating contribution, as argued in \cite{Kostelecky02}.   The  CPT theorem  states that CPT symmetry arises under assumptions through the combination of Lorentz-symmetry and quantum-mechanical assumptions. Then, if CPT symmetry is broken, one or more of the assumptions necessary to prove the CPT theorem should not be valid. Once both Lorentz  and CPT invariance involve spacetime transformations, it is natural to suspect that CPT violation implies Lorentz-symmetry breakdown. This  statement was  rigorously proven  in  \cite{Greenberg02, Greenberg06, Lehnert16}. However, Lorentz-symmetry breaking does not necessarily imply CPT violation.
To obtain the fermion propagator, it is convenient to define the matrix $D=p_\mu \gamma^\mu +d_\mu\gamma^\mu \gamma_5 - m\,\mathds{1}$ that acts on the  $u(p)$ amplitude in Eq. \eqref{eqD1}, where $d^\mu = (\vartheta \, \bm{p}^2,{\bf 0})$ is a time-like vector whose the component depends on the ${\bm p}^2$. Thereby, the fermion propagator is the inverse of the $D$-matrix:
\begin{widetext}
\begin{align} \label{propagador}
D^{-1} =&\; \Big[ (p^2-m^2 -d^2)\left(m+p_\mu \gamma^\mu +d_\mu\gamma^\mu \gamma_5 \right)  -2(p\cdot d )(d_\mu \gamma^\mu - p_\mu \gamma^\mu \gamma_5) -2m\,(\,p^\alpha \, d^\beta -p^\beta \, d^\alpha\,)\, \varepsilon_{\alpha\beta\mu\nu} \, \Sigma^{\mu\nu}  \Big] \, \Delta^{-1} \; ,
\end{align}
\end{widetext}
where the denominator $\Delta$ is given by
\begin{align} \label{RDf}
\Delta =&\; (p^2-m^2)^2-4(p\cdot d)^2 +2d^2(p^2+m^2) + d^4 \; .
\end{align}
The dispersion relation of the particle is the propagator's pole at $\Delta=0$. Restricting these results to the case of the time-like $4$-vector $d^\mu=(\vartheta \, \bm{p}^2,{\bf 0})$, the fermion dispersion relation is reduced to
\begin{align}
4 p^2 \vartheta^2 \bm{p}^4 +\left(p^2-m^2- \vartheta^2 \bm{p}^4 +2 E \vartheta \bm{p}^2 \right)
\nonumber\\
\times\left( p^2-m^2- \vartheta^2 \bm{p}^4 -2 E \vartheta \bm{p}^2 \right)
 =0 \; . \label{RDLQG}
\end{align}
The corresponding energy solutions of Eq. (\ref{RDLQG}) are the positive, $+E_{\pm}({\bf p})$, and negative, $-E_{\pm}({\bf p})$, energy eigenstates, 
where
\begin{align} \label{RDEnergy}
E_{\pm}({\bf p}) = \sqrt{\bm{p}^2+m^2+ \vartheta^2 \, \bm{p}^4 \pm 2 \vartheta (\bm{p}^2)^{3/2} }  \; ,
\end{align}
that confirm the results obtained in \cite{Li23}. 
In the rest frame of the fermion, we have no mass splitting
\begin{align}
E_0 = m.
\end{align}
This indicates that the ($\pm$) signals in Eq. \eqref{RDEnergy} correspond to two energy eigenstates of the same fermion.
The $i$-th component of the group velocity becomes
\begin{align}
\bm{v}_i = \dfrac{\partial E}{\partial \bm{p}_i} \; ,
\end{align}
splitting into time and space parts, and also taking the  total differential of Eq. (\ref{RDLQG}), one can show the group velocity of the propagating wave is
\begin{align}
 \bm{v}_i =  \left[\dfrac{6\,(E^2-m^2-\bm{p}^2)}{E^2-m^2-\bm{p}^2- \vartheta^2\,\bm{p}^4} +  2\, \vartheta^2\,\bm{p}^2-5 \right] \dfrac{\bm{p}_i}{E} \; ,
\end{align}
which reduces to the usual case, $\bm{p}_i/E$, whether we turn off the LSV terms. Substituting the positive energy solutions, one obtain
\begin{eqnarray} \label{gvel}
\bm{v}^{\pm}_i=  \left[ \frac{1+\,  2  \, \vartheta^2\,{\bf p}^2 \, \pm \, 3  \vartheta \, |{\bf p}|}{\sqrt{\bm{p}^2+m^2+ \vartheta^2 \, \bm{p}^4 \pm 2\vartheta (\bm{p}^2)^{3/2} }} \right]  \bm{p}_i \; .
\end{eqnarray}
The result with a $(+)$ sign on the right-hand side of Eq. \eqref{gvel} is superluminal, $|\bm v_+| > 1$, consistent with the phenomenology discussed in \cite{Li23}. At this point, it is important to emphasize that we are dealing with a possible break of Lorentz symmetry, and for this reason, the familiar causal structure of special relativity may not be such a rigid rule  that must hold. 
For the purposes of Section \ref{sec5}, it is interesting to present the form of the dispersion relation in the high momentum limit at leading order two,
\begin{align} 
E_{\pm}({\bf p}) = |\bm{p}| + \dfrac{m^2}{|\bm{p}|} \pm \vartheta |\bm{p}|^2 \; ,
\end{align}
followed by the group velocity modulus associated with this regime,
\begin{align} \label{groupvel}
|\bm{v}^{\pm}| = 1 - \dfrac{m^2}{2|\bm{p}|^2} \pm 2 \vartheta |\bm{p}|, 
\end{align}
which has the same interpretation regarding the causal aspects of Eq. \eqref{gvel}.
\section{The positive energy solutions for the modified Dirac equation} \label{sec3}
From the energy eigenstates, we can obtain the positive energy solutions for the modified Dirac equation in the
laboratory reference frame. We consider the following plane wave solution for a free fermion $\psi$:
\begin{align} \label{solmom}
   \psi (x, \pm s)
   = u(p,\pm s) \, e^{-ip \cdot x}
   = \begin{pmatrix}
   u_a(p,\pm s) \\
    u_b(p,\pm s)
\end{pmatrix}    e^{-ip\cdot x} \; ,
\end{align}
in which the positive energy is described by the four-momentum fermion
$p^{\mu} = \left(E_{\pm}, \bm{p} \right)$, in a laboratory reference frame, and with spin projection $(\pm s)$.
Using this definition, the equation field is
\begin{align} \label{waveeq}
\left( E_\pm \gamma^0 -p_i \gamma^i - m  +\vartheta \bm{p}^2 \gamma^0 \gamma_5 \right)  u(p,\pm s) =0 \; .
\end{align}
In matrix form, this equation leads to the coupled system
\begin{subequations}
\begin{eqnarray}
&\!\!\!\!\!\!\!\!
\left(E_\pm   \! - \! m \right) \! u_a(p,\pm s) \! +  \! \left(\vartheta \bm{p}^2 \mathds{1} \! - \! \bm{\sigma} \! \cdot \!  \bm{p}  \right) \! u_b(p,\pm s) \! = \! 0 \; ,
\\
&\!\!\!\!\!\!\!\!
\left(E_\pm  \! + \!  m \right) \! u_b(p,\pm s) \! + \! \left(\vartheta \bm{p}^2 \mathds{1} \! - \! \bm{\sigma} \! \cdot \! \bm{p}   \right) \! u_a(p,\pm s) \! = \! 0 \; ,
\end{eqnarray}
\end{subequations}
where $\mathds{1}$ is the $2\times 2$ identity. Thus, we obtain the relation
\begin{align}\label{relationubua}
 u_b(p,\pm s) = \dfrac{\bm{\sigma}\cdot\bm{p}-\vartheta \bm{p}^2\,\mathds{1}}{E_\pm + m} \, u_a(p,\pm s) \; .
\end{align}
Therefore, the general solution for a positive energy eigenstate in momentum space is written as
\begin{align}
u(p,\pm s)  = \begin{pmatrix}
\mathds{1} \\
\dfrac{\bm{\sigma}\cdot\bm{p}-\vartheta \,\bm{p}^2\, \mathds{1}  }{E_\pm +m}
\end{pmatrix} u_a(p,\pm s) \; .
\label{Esp1}
\end{align}
As in the case of the usual Dirac equation, we construct the spin up solution from the amplitude
\begin{align}
u_a(p,+ s)  = N_+ \begin{pmatrix}
  1 \\ 0
\end{pmatrix} ,
\end{align}
that substituting in Eq. (\ref{relationubua}), the $u_b$-amplitude for $(+s)$ is
\begin{align}
u_b(p,+ s) &= \dfrac{N_+}{E_\pm + m}  \begin{pmatrix}
p_z -\vartheta \bm{p}^2 \\ p_x +i p_y
\end{pmatrix} \; .
\end{align}
Thereby, the general solution has the form
\begin{align} \label{sol22}
u(p,+ s) = N_+ \begin{pmatrix}
  1 \\0 \\ \dfrac{p_z -\vartheta \bm{p}^2}{E_\pm + m} \\ \dfrac{ p_x +i p_y }{E_\pm + m}
\end{pmatrix} \; ,
\end{align}
where $N_+$ is a normalization constant. Using the normalization condition $\bar{u}(p,+s) \, u(p,+ s)= 2m $,
the $N_{+}$-constant is
\begin{align}
N_+
= \sqrt{\dfrac{2m(E_\pm +m )^2}{2m(E_\pm + m) +\vartheta (2p_z \pm |\bm p|) \bm p^2 }}\; .
\end{align}
The solution describing spin up is
\begin{align}\label{psi+s}
\psi (x, + s)=&
\sqrt{\dfrac{2m(E_\pm +m )^2}{2m(E_\pm + m) +\vartheta (2p_z \pm |\bm p|) \bm p^2 }}
\nonumber \\
&\times
\begin{pmatrix}
  1 \\0 \\ \dfrac{p_z -\vartheta \bm{p}^2}{E_\pm + m} \\ \dfrac{ p_x +i p_y }{E_\pm + m}
\end{pmatrix} e^{-i p \cdot x}
\, .
\end{align}
For the case of the spin-down solution, the $u_a$-amplitude becomes
\begin{align}
u_a(p,-s)  = N_- \begin{pmatrix}
  0 \\ 1
\end{pmatrix} ,
\end{align}
and thus, we obtain
\begin{align}
u_b(p,-s) &=    \dfrac{N_-}{E_\pm + m}
\begin{pmatrix}
p_x- i p_y \\ -p_z - \vartheta \bm{p}^2
\end{pmatrix} \; .
\end{align}
The general solution for $(-s)$ has the form
\begin{align} \label{sol27}
u(p,- s) = N_{-} \begin{pmatrix}
  0 \\1 \\ \dfrac{ p_x- i p_y }{E_\pm + m} \\ \dfrac{ -p_z - \vartheta \bm{p}^2 }{E_\pm + m}
\end{pmatrix} ,
\end{align}
where $N_{-}$ is the correspondent normalization constant determined by the condition
$\bar{u}(p,-s)\, u(p,-s)= 2m$. Thereby, we obtain
\begin{align}
N_{-}
= \sqrt{\dfrac{2m(E_\pm +m )^2}{2m(E_\pm + m) -\vartheta (2p_z \mp |\bm p|) \bm p^2 }}\; ,
\end{align}
and the correspondent solution is given by
\begin{align}\label{psi-s}
\psi (x, -s)=&  \sqrt{\dfrac{2m(E_\pm +m )^2}{2m(E_\pm + m) -\vartheta (2p_z \mp |\bm p|) \bm p^2 }}
\nonumber \\
&\times \begin{pmatrix}
  0 \\1 \\ \dfrac{ p_x- i p_y }{E_\pm + m} \\ \dfrac{ -p_z - \vartheta \bm{p}^2 }{E_\pm + m}
\end{pmatrix}  e^{-i p \cdot x}
\, .
\end{align}
The limit $\vartheta \rightarrow 0$ recovers all known results of the usual Dirac equation.
Notice that, the normalization constants are not equal by the introduction of $\vartheta$-scale,
{\it i.e.}, $N_{+} \neq N_{-}$. Consequently, the introduction of the LQG-scale breaks the
symmetry in the Dirac equation solutions, and still includes degenerate solutions $(\pm)$ in Eqs.
(\ref{psi+s}) and (\ref{psi-s}), respectively.
\section{The $U(1)$ conserved current} \label{sec4}
Consider $\widehat{{\cal O}}$ a operator that acts of generic mathematical objects $A$ and $B$, such as scalars, vectors and spinors. Throughout the text, we assume the definitions $A \overleftarrow{\widehat{{\cal O}}} B = (\widehat{{\cal O}} A) B$ and $ A \overleftrightarrow{\widehat{{\cal O}}} B \equiv   (\widehat{{\cal O}} A) B -  A(\widehat{{\cal O}} B)$. Then, assuming that the transform under global $U(1)$ for the matter field $(\psi)$
\begin{align}
\psi \, \longmapsto \, \psi^{\prime} = e^{-i\alpha} \, \psi \; ,
\end{align}
the fermion action in Eq. (\ref{ActionDirac1}) is invariant. On the other hand, if we introduce a local $U(1)$ transformation for an infinitesimal $\alpha$-parameter, the fermion action in Eq. \eqref{ActionDirac1} at first order in $\alpha$ is
\begin{align} \label{dfrrg}
    \mathcal{S} \, \longmapsto \, \mathcal{S}^{\prime}
      =&\;  \mathcal{S}  - \int    d^4x \, (\partial_ \mu \alpha) \, \bar{\psi} \gamma^\mu \psi
      \nonumber \\
       &\; + i \vartheta  \int   d^4x \, (\bm{\nabla} \alpha)  \cdot   \bar{\psi} \gamma^0 \gamma_5 \vecev{ \bm{\nabla}} \psi \; .
\end{align}
The variation of Eq. \eqref{dfrrg} with respect to $\alpha$-parameter yields the $4$-current
\begin{align} \label{currentU1}
J^\mu = (J^0, J^i) =\Big(\bar{\psi}\gamma^0\psi, \bar{\psi}\gamma^i \psi - i\vartheta \bar{\psi} \gamma^0 \gamma_5 \vecev{\partial}^{i} \psi  \Big) \; ,
\end{align}
that satisfies the continuity equation $\partial_\mu J^\mu =0$. The minimal coupling with the electromagnetic field is introduced substituting the derivative operator by the covariant derivative operator $\partial_\mu \longmapsto D_{\mu}=\partial_\mu +ie \, A_\mu$. The new action for
the QED is
 \begin{align} \label{QEDaction}
 S = &\; \int   d^4x   \bigg[ -\dfrac{1}{4}\,F_{\mu\nu}F^{\mu\nu}  - eA_\mu\bar{\psi} \gamma^\mu \psi \nonumber \\
 &\; - i\vartheta e \bm{A} \cdot \bar{\psi} \gamma^0 \gamma_5 \vecev{ \bm{\nabla}} \psi  + \vartheta e^2 \bm{A}^2 \, \bar{\psi}\gamma^0\gamma_5 \psi  \nonumber \\
 &\; + \bar{\psi} \left(i\gamma^\mu\partial_\mu -\vartheta \, \gamma^0 \gamma_5\,\bm \nabla^2 - m\right)\psi \bigg] \, ,
 \end{align}
which includes a new quartic coupling involving the quadratic vector potential with the fermions fields.

We should here open up a parenthesis to state our position why we do not include LQG contributions in the photon sector. As it is clear from the Eq. \eqref{QEDaction} above, the photon is described by the usual Maxwell kinetic term. Our main purpose in this work is to keep track of the influence of the LQG effects on the electromagnetic properties of charged fermions. So, in a first moment, we take the photon free from the LSV corrections and try to see how this sort of ``plain'' photon probes the LQG-corrected electromagnetic current of fermionic matter. We however consider relevant - though it is not what we are doing here - to investigate the full-fledged QED action, with both the photon and fermion sectors including the LQG contributions.

Since the Bianchi identity is the usual expression, one can verify that magnetic Gauss and Faraday-Lenz laws are unchanged. Also the electric Gauss law is protected, as long as the new current does not affect the electrical charge density,
\begin{align}
\bm \nabla \cdot \bm E = -e \, \psi^{\dagger} \, \psi \; .
\end{align}
Just the Ampère-Maxwell law is modified by the spatial current density
\begin{equation} \label{ampere}
\bm{ \nabla} \times \bm B = \partial_t \bm E \! + \! e  \bar{\psi} \bm \gamma \psi
\! - \! 2 \vartheta e^2 \bm A \bar{\psi}\gamma^0 \gamma_5 \psi
\! +\! i e  \vartheta  \bar{\psi} \gamma^0 \gamma_5 \vecev{ \bm{\nabla}} \psi \; .
\end{equation}
The displacement current on the right-hand side of Eq. \eqref{ampere} coincides with the spatial part of $U(1)$ current in Eq. (\ref{currentU1}). Thereby, we can conclude that the conserved $U(1)$ and the electromagnetic current is the same in this approach.  
Substituting the solutions given by Eq.  \eqref{Esp1} in the spatial current density of  Eq. \eqref{currentU1}, we obtain the result in the momentum space
\begin{align}
     \bm J = &\;   \dfrac{1}{E_\pm +m } \, u^\dagger_a(p')  \Bigg\{\bigg[ 1-  \dfrac{\vartheta^2}{2} (\bm{q}^2+\bm{\ell}^2) \bigg] \bm \ell   - (\bm q \times \bm \sigma)  \nonumber \\
     &\; -\dfrac{\vartheta}{2} \, \bm{q}^2 \, \bm \sigma  -  \vartheta\bigg(\dfrac{\bm{\ell}^2}{2} \, \delta_{ij} - \bm \ell_i \, \bm \ell_j \bigg) \bm  \sigma_j \Bigg\} u_a(p) \; ,
    \label{result10}
\end{align}
where we have omitted the spins projections $(\pm s)$, and also have defined the photon's transfer 3-momentum $\bm q = \bm p' - \bm p$, and the total  3-momentum $\bm \ell = \bm p + \bm p'$. Since the term proportional to $\vartheta^2$ is too small, the only contribution of a granular space-time in the current comes from a helicity-type interaction with respect to the transfer momentum $\bm q$, and a projection of the spin on $\bm \ell$-direction.

\section{The Gordon decomposition} \label{sec5}
\begin{figure}[H]
\centering
\includegraphics[scale=0.65]{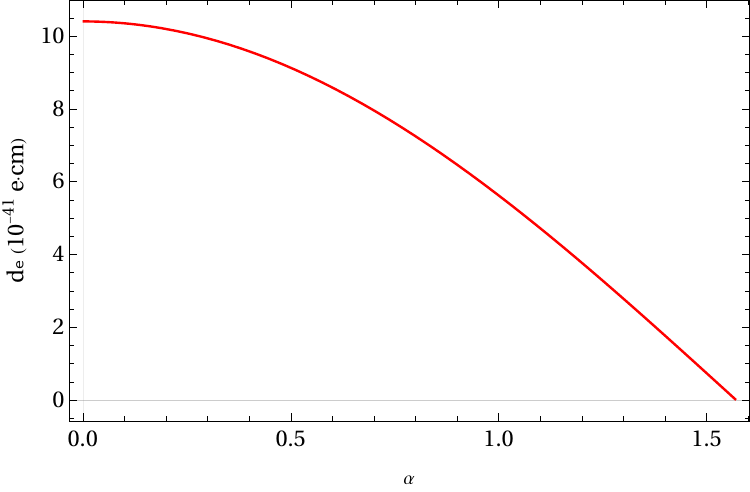}
\caption{Upper bounds on the $d_{e}$ as function of the $\alpha$-angle. In this plot, we fix the momenta 
and the energy at the MeV QED scale : $| \bm p | = | \bm  p^{\prime} | = q_0 = 1 $ MeV. }
\label{fig1}
\end{figure}

In this Section, we introduce the Gordon decomposition procedure for the modified Dirac equation, in which the interpretation of the current coupled to the electromagnetic field is an important key to understand the form factors, as well as the electron's magnetic and electric dipole momentum. In order, we must emphasize that the conserved current of the model given by Eq. \eqref{currentU1} presents the interesting fact that its temporal and spatial components are distinct in the sense that the former does not undergo the effects of LSV, while the latter does. This fact suggests that the usual procedure for deducing Gordon's identity can no longer be applied at least to the spatial part of Eq. \eqref{currentU1}, which differs from the usual case by the LSV terms. Thereby, we have two Gordon identities, the first for the time component of Eq. \eqref{currentU1}, and the second one for the spatial part. For the current's time-component, we go back to the Eq. \eqref{Dirac1}, whose Dirac conjugate equation for $\bar{\psi}(x)$ is given by Eq. \eqref{eqAdjoint}, and combining Eq. \eqref{Dirac1} with Eq. \eqref{eqAdjoint}, we multiply by $\gamma^0 \, \psi(x)$ at the right-hand side of Eq. \eqref{eqAdjoint}, and also multiply by $\bar{\psi}(x) \, \gamma^0$ at the left-hand side. By subtracting these equations, we get
\begin{align}\label{j0config}
J^0= - \dfrac{i}{2m}  \bar{\psi} \vecev{\partial}_t \psi- \dfrac{1}{m} \partial_{\mu} (  \bar{\psi} \Sigma^{\mu 0} \psi )
- \dfrac{\vartheta}{2m}  \bar{\psi} \gamma_5  \vecev{\bm{\nabla}}^2 \psi \; ,
\end{align}
which is the Gordon identity for $J^0$-component of the conserved current.
For the spatial part of the conserved current, we repeat the procedure above by multiplying by ${\bm \gamma} \, \psi + i\vartheta \, (\cev{\bm{\nabla}})\gamma^0\gamma_5\psi -i\vartheta\,\gamma^0\gamma_5(\bm{\nabla}\psi)$ on the right-hand side of Eq. \eqref{eqAdjoint}, and also multiplying by $\bar{\psi} \, {\bm \gamma} +i \vartheta \, (\bm{\nabla}\bar{\psi})\gamma^0\gamma_5 -i\vartheta \, \bar{\psi}\gamma^0\gamma_5\bm{\nabla}$ the left-hand side, with the LSV terms being necessary due to the new structure of the current. Therefore, the Gordon identity for $J^{i}$ becomes as given below:
\begin{eqnarray} \label{jiconfig}
J^{i} \!&=&\! \dfrac{i}{2m} \, \bar{\psi} \partial^{i} \psi - \dfrac{1}{m} \, \partial_\mu ( \bar{\psi}\,\Sigma^{\mu i } \psi)
+\dfrac{\vartheta }{2m} \, (\partial_t  \bar{\psi}) \gamma_5 \partial^{i} \psi
\nonumber \\
&&
\hspace{-0.5cm}
-\dfrac{2i\vartheta }{m} \, ( \bm \nabla \bar{\psi}) \, \Sigma^{0i }\gamma_5 \cdot (\bm{\nabla}\psi)  +\dfrac{i \vartheta^2}{2m} \, \partial^{i}  (\bar{\psi} \vecev{\bm\nabla}^2 \psi) \; .
 \end{eqnarray}
The structure of both the components of the current as given in Eqs. \eqref{j0config} and \eqref{jiconfig} shall be applied in the next Section, where we shall be calculating the interparticle electron-electron ($e^{-}e^-$) non-relativistic potential mediated by a photon exchange by means of an elastic scattering.  The role of the current components will become clearer once the expression for the $e^-e^-$ potential is written in terms of its velocity and spin content.    
From the action in Eq. \eqref{QEDaction}, the three-vertex coupling of the leptons current with the electromagnetic (EM) field is described by
\begin{align}
S_{int}^{(3)} = -e\int d^4 x \Big[ \, \bar{\psi} \gamma^{\mu}  \psi  A_{\mu} + i\vartheta  \bm{A} \cdot \bar{\psi} \gamma^0 \gamma_5 \vecev{ \bm{\nabla}} \psi \, \Big] \; ,
\end{align}
which, in the momentum space, reads as
\begin{align}
S_{int}^{(3)} =&\; -e \int \dfrac{d^4 p'}{(2\pi)^4} \frac{d^4 p}{(2\pi)^4}
\Big[ \, \bar{u}(p') \gamma^{\mu} u(p)   A_{\mu}(q)
\nonumber \\
&\; -\vartheta \, \bar{u}(p') \gamma^{0} \gamma_5 u(p)    (\bm p'+ \bm p) \cdot \bm A (q) \, \Big] \; .
\end{align}
Using the Gordon identities, the time-component for the coupling is
\begin{align}\label{baruA0u}
\bar{u} & (p')
 \gamma^0  u(p)A_0 (q) =
\nonumber \\
& =  \bar{u}(p') \bigg[ \, \dfrac{\ell^0}{2m} \! - \! \dfrac{iq_{\mu}}{m} \, \Sigma^{\mu 0}  \! -  \vartheta \, \dfrac{\bm \ell \cdot \bm q}{2m} \gamma_5 \, \bigg] u(p) A_0 (q) \; ,
\end{align}
where $q^{\mu} = p^{\prime\mu} - p^{\mu}$ and $\ell = p^{\prime\mu} + p^{\mu} $ are the photon's transfer  and total $4$-momentums, respectively. This coupling is interpreted as the interaction energy between the charge density and the scalar potential $A^{0}\equiv V$. In  configuration  space, the   $\vartheta$-term in Eq. (\ref{baruA0u}) for the electrostatic case becomes
\begin{eqnarray}
-\vartheta \, \dfrac{\bm \ell \cdot \bm q}{2m} \, V \, \gamma_5 =-\frac{i\vartheta}{2m} (\bm \ell \cdot \bm E) \, \gamma_{5} \; ,
\end{eqnarray}
where $\bm q \rightarrow -i\bm \nabla$, and ${\bf E}=-\bm \nabla V$. Thus, the electric dipole momentum emerges with $\bm \ell \cdot \bm E$,
and the ratio $(\vartheta/2m)$ modulates the LQG contribution to the electron's electric dipole momentum.
From the spatial part, the current density is
\begin{align} \label{DGespacial}
\bar{u}(p') \big[  & {\bm \gamma}   -  \vartheta  \bm \ell \gamma^{0} \gamma_5 \big] u(p)  \cdot  \bm A (q) = \nonumber \\
=&\; \bar{u}(p') \bigg[    \dfrac{\bm \ell}{2m}  -  \dfrac{iq_\mu}{m} \Sigma^{\mu i} +  \dfrac{\vartheta q^0 \bm \ell}{2m} \gamma_5 \nonumber \\
&\; -  \dfrac{i\vartheta (\bm \ell^2 \! - \! \bm q^2)}{2m} \Sigma^{0i}\gamma_5  \!- \! \dfrac{\vartheta^2 (\bm q \!\cdot\! \bm \ell)\bm q}{2m}\bigg] u(p) \! \cdot \! \bm A (q) \; .
\end{align}
In coordinate space, the last term can be eliminated by the Coulomb gauge, $\bm \nabla \cdot {\bf A}=0$.
An important aspect arising from the Gordon decomposition in Eq. \eqref{DGespacial} is the contribution to the electron's electric dipole moment (EDM). Specifically, it is associated with the term proportional to $\Sigma^{0i}\gamma_5$. This term defines an interaction action for the EDM in momentum space such that
\begin{align}
{\cal S}_{\textrm{\tiny EDM}} =  \int \dfrac{d^4 p'}{(2\pi)^4} \frac{d^4 p}{(2\pi)^4} \, \bar{u}(p') \, d_e \, \bm \Sigma \cdot \bm E (q) \, u(p) \; ,
\end{align}
where ${\bf E}(q)$ is the momentum space version of the electric field,
\begin{align}
\bm \Sigma = \begin{pmatrix}
    \bm \sigma & 0 \\
    0 & \bm \sigma
\end{pmatrix} \; ,
\end{align}
and the EDM is
\begin{align} \label{EDMde}
d_e = \dfrac{\vartheta e }{8m} \left( \dfrac{\bm \ell^2 \! - \! \bm q^2}{q_0} \right)= \dfrac{\vartheta e }{2m} \left( \dfrac{\bm p' \cdot \bm p}{q_0} \right) \; .
\end{align}
The corresponding Lagrangian density is 
\begin{align}
\mathcal{L}_{\textrm{\tiny EDM}} = \bar{u}(p') \, d_e \, \bm \Sigma \cdot \bm E (q) \, u(p) \; .
\end{align}
The most recent experimental results for the EDM, obtained by using electrons confined inside molecular ions subjected to a huge intramolecular electric field, provide an upper bound of $d_e < 4.1 \times 10^{-30}$ $e \, \cdot $ cm \cite{JILA23}. We can use this result to estimate a bound on the $\vartheta$-parameter. In order, the Eq. \eqref{EDMde} can be written as
\begin{align}
d_{e}= \frac{\vartheta\,e}{2m} \left( \dfrac{| \bm p | | \bm p^{\prime} | \cos\alpha }{q_0} \right) \leq 4.1 \times 10^{-30} \,\, \mbox{e} \cdot \mbox{cm} \; ,
\end{align}
where $\alpha$ is the angle between $\bm p$ and $\bm p^{\prime}$. From the Eqs. \eqref{Dirac1} and \eqref{Coperator}, the $\vartheta$-parameter is defined by
\begin{eqnarray}
\vartheta=\frac{\kappa_7}{4{\cal L}}\left( \frac{\ell_{P}}{{\cal L}} \right)^{\!\! \Upsilon_{\!f}}\ell_{P}^2 \; .
\end{eqnarray}
In Ref. \cite{Li23}, it is argued that $\Upsilon_{f}=-1$ and $\kappa_7=37.5$ for neutrinos. Here, it is necessary to carry on a similar analysis to determine the values of these parameters, when dealing with electrons. To follow this path, we base our analysis on the data from Ref. \cite{Li22a}, which studies the LSV effects in the propagation velocity of a charged lepton, not just in neutrinos and photons, as it is the case in Ref. \cite{Li23}. Following the steps described in Ref. \cite{Li22a}, we can express the modulus of our group velocity, Eq. \eqref{groupvel}, in the form
\begin{align}
    |\bm v_{\pm}| = 1 - \dfrac{m^2}{2\bm p^2} \pm  \dfrac{|\bm p|}{E_{\tiny \textrm{LSV}}} , 
\end{align} 
which, to match the general results presented in Ref. \cite{Li22a}, requires that $\Upsilon_f = -1$ and, consequently,
\begin{align}
    \dfrac{\kappa_7 \ell_P}{2} = \dfrac{1}{E_{\tiny \textrm{LSV}}}, 
\end{align}
with $E_{\tiny \textrm{LSV}} \gtrsim 9.4 \times 10^{25}$ G$e$V for the highest-energy event from the Crab Nebula \cite{LHAASO21}. All of this implies that $\kappa_7 \approx 2.6 \times 10^{-7}$. Then, considering a energy scale at $| \bm p | = | \bm  p^{\prime} | = q_0 = 1 $ MeV, and the electron mass of $m=0.5$ MeV, we can illustrate our $d_e$ values  as a function of the $\alpha$-angle in the plot of the FIG. \ref{fig1}. Since all $d_e$ values in FIG. \ref{fig1} are on the order of $10^{-40}$ to  $10^{-41} \; e  \cdot $cm, the EDM predicted by this LQG model falls within the current experimental limit of $d_e < 4.1 \times 10^{-30}\; e  \cdot $cm \cite{JILA23}. 

It is also important to emphasize that this result is consistent with the Standard-Model prediction for the EDM  of charged leptons using hadron effective models as given by \cite{Yamaguchi20}:  $d_e = 5.8 \times 10^{-40}$ $e\cdot$cm, with an error bar of $70\%$, exceeding the conventionally known four-loop level elementary contribution by several orders of magnitude.

\section{Electromagnetic Interaction Between Fermions} \label{sec6}
\begin{figure}[H]
    \centering
    \includegraphics[scale=1.5]{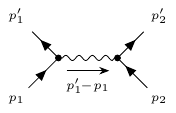}
    \caption{An $e^- e^-$ tree-level elastic  scattering involves incoming $4$-momenta $p_1$ and $p_2$, and outgoing $4$-momenta $p'_1$ and $p'_2$, mediated by a photon that transfers $4$-momentum $p'_1 - p_1 = -(p'_2 - p_2)$ during the process. }
    \label{diagram}
\end{figure}

In this Section, we aim to present the implications of the electromagnetic current Gordon decomposition structure previously shown by examining its influence on the Coulomb potential. To achieve this, we analyze the interaction between two fermions, as described by the solutions in Eqs. \eqref{sol22} and \eqref{sol27}, in a tree-level elastic scattering considering the possibility of spin-flip. The amplitude of this process can be characterized by the Feynman diagram depicted in FIG. \ref{diagram}.
In this tree-level elastic scattering, the two fermions, each with charges $e_1$ and $e_2$, possess incoming $4$-momenta denoted as ${p_1}^\mu=(E_1,\bm p_1)$ and ${p_2}^\mu= (E_2,\bm p_2)$, and outgoing $4$-momenta denoted as ${p'_1}^\mu=(E'_1,\bm p'_1)$ and ${p'_2}^\mu=(E'_2,\bm p'_2)$, respectively. Furthermore, it is convenient for us to work in the center of mass  reference frame, with the parameterization in terms of the following independent momenta: the transferred momentum, $q=p'_1 -p_1 = -(p'_2-p_2)$, and the average momentum, $\frac{\ell}{2} = \frac{(p'_1 +p_1)}{2} = \frac{(p'_2 +p_2)}{2} $. Through the conservation of energy and momentum, we have that $q^0 = 0$ and $\bm q \cdot \bm \ell = 0 $ to simplify the amplitude of FIG. \ref{diagram}.
In order to include spin and velocity corrections to the potential, we shall employ the method described in \cite{Moody84, Dobrescu06}. The prescription is that the potential,  $V$, can be obtained from the first Born approximation, i.e., by performing the Fourier integral of the non-relativistic  (NR) amplitude, $\mathcal{M}_{\tiny{\textrm{NR}}}$, with respect to the transferred momentum $\bm q$,
\begin{align}
    V(\bm x) = - \int \dfrac{d^3  \bm q}{(2\pi)^3} \mathcal{M}_{ \tiny{\textrm{NR}}} e^{i\bm q \cdot \bm x} , 
\end{align}
where $\mathcal{M}_{\tiny{\textrm{NR}}}$ is related to the relativistic Feynman amplitude, $\mathcal{M}$, according to the conventions of \cite{Maggiore05}, through
\begin{align}
 \mathcal{M}_{ \tiny{\textrm{NR}}} = \dfrac{1}{\sqrt{2E_1}}\dfrac{1}{\sqrt{2E'_1}} \dfrac{1}{\sqrt{2E_2}} \dfrac{1}{\sqrt{2E'_2}}    \mathcal{M}.
\end{align}
The amplitude $\mathcal{M}$ is obtained by applying Feynman rules, incorporating the fields, interaction vertices, and propagators. In the case to be considered here, it can be rewritten as $\mathcal{M} \sim J_{(1)} \langle \textrm{prop} \rangle J_{(2)}$,  where  $ J_{(1)}$ and $ J_{(2)}$ represent the sources/currents, and $\langle \textrm{prop} \rangle $ denotes the propagator, with the possible Lorentz indices of the representations being omitted. Therefore, taking the photon propagator in the Feynman gauge,
\begin{align}
    \langle A_\mu A_\nu \rangle =i \dfrac{\eta_{\mu\nu}}{q^2},
\end{align}
for the process in FIG. \ref{diagram} we have the amplitude
\begin{align}
   \!\!\!\!\!  i\mathcal{M} &= -e_1e_2 J^{\mu}_{(1)}  \langle A_\mu A_\nu \rangle J^{\mu}_{(2)} \nonumber \\
    &=-e_1e_2 \bar{u}(p'_1) \gamma^\mu u(p_1)  \langle A_\mu A_\nu \rangle \bar{u}(p'_2) \gamma^\nu u(p_2)  ,
\end{align}
with $J^\mu $ here denoting the vector current of the model given by Eq. \eqref{currentU1} in the momentum space. Then, utilizing the continuity equation $q_\mu J^\mu =0$ and $q^0 =0$, in the NR limit yields
\begin{align}
     \mathcal{M}_{ \tiny{\textrm{NR}}} &=  - \dfrac{e_1e_2}{\bm q^2}\dfrac{J^\mu_{(1)}J_{(2)\mu}}{4 E_1 E_2}   \nonumber \\
     &= - \dfrac{e_1e_2}{\bm q^2}\dfrac{J^0_{(1)}J^0_{(2)}- \bm J_{(1)} \cdot \bm J_{(2)} }{4 E_1 E_2}  .
\end{align}
To proceed, we must express both $J^0_{(1)}$ and $\bm J_{(1)}$ for the particles with a subscript in terms of their respective Gordon decompositions, Eqs. \eqref{baruA0u} and \eqref{DGespacial}, and positive-energy solutions, Eqs. \eqref{sol22} and \eqref{sol27}, in the NR limit. The NR limit of the solution is given by
\begin{align} \label{solNR1}
    u_{\textrm{\tiny NR}}(p_1) \approx  \sqrt{2m}
\begin{pmatrix}
\chi_1 \\[0.2cm]
\dfrac{\bm{\sigma}  \cdot \bm{p}_1 -\vartheta\bm p^2_1 }{2m} \chi_1 
\end{pmatrix} 
\end{align}
and its Dirac's conjugate
\begin{align}\label{solNR2}
    \bar{u}_{\textrm{\tiny NR}}(p'_1) \approx  \sqrt{2m}\left(  {\chi'_1}^\dagger, {\chi'_1}^\dagger \dfrac{\bm{\sigma}  \cdot \bm{p}'_1 -\vartheta {\bm p'_1}^2 }{2m}  \right)  \gamma^0.
\end{align}
The spin-up and -down solutions of Eq. \eqref{solNR1} can be obtained, respectively, from the spin-up and -down solutions in the rest frame, which are respectively given by

\begin{align}
    \chi_1 = \begin{pmatrix}
        1 \\ 0
    \end{pmatrix} \;\; \textrm{and} \;\;    \chi_1 = \begin{pmatrix}
        0 \\ 1
    \end{pmatrix} .
\end{align}
As a consequence, the spin-up and -down solutions for Eq. \eqref{solNR2}, respectively, are given by
\begin{align}
   {\chi'_1}^\dagger = \left(1,0\right) \;\; \textrm{and} \;\;  {\chi'_1}^\dagger = \left(0,1\right) .
\end{align}
Therefore, with all these statements and by imposing the conditions of elastic scattering, one can write
\begin{align} \label{J10}
    J^0_{(1)}  \! = \!  \left[    2m_1 \! + \! \dfrac{1}{8m_1} (\bm \ell^2 \! + \! 7 \bm q^2)   \right] \! \delta_1 \!  + \! \dfrac{3}{4m_1} i (\bm q \! \times \! \bm \ell) \! \cdot \! \langle \bm S_1 \rangle,
\end{align}
and
 \begin{align} \label{J1bm}
   \bm J_{(1)}  \! = \!  \left[  1 \! + \! \dfrac{(\bm q^2 \! - \! \bm \ell^2)}{16m_1^2}  \right] \! \bm \ell \delta_1 \!   - \! 2i  \bm q \! \times \! \langle {\bm  S_{1} \rangle} \! + \! \vartheta (\bm \ell^2 \! - \! \bm q^2)  \langle {\bm  S_{1}} \rangle ,
\end{align}
 where we defined the quantities 
 \begin{align}
 &{\chi'_1}^\dagger \chi_1 = \delta_1 , \\
 &{\chi'_1}^\dagger \dfrac{\bm \sigma}{2} \chi_1  = \langle \bm S_1 \rangle .
\end{align}
To express $J^0_{(2)}$ and $\bm J_{(2)}$, one simply needs to replace the index $1$ by $2$ in Eqs. \eqref{J10} and \eqref{J1bm}, as well as replace the momenta $\bm q$ and $\bm \ell$ by $-\bm q$ and $-\bm \ell$, respectively. With these results, and still in the NR limit, one can write the expression for the interparticle potential in the form 
\begin{widetext}
  \begin{align} \label{potentialV}
    V (\bm x) =&\;  e_1e_2 \bigg\{ \left[ \dfrac{5}{16} \left(\dfrac{1}{m_1^2} + \dfrac{1}{m_2^2} \right) \delta_1 \delta_2  - \dfrac{2}{3} \dfrac{\langle {\bm  S_{1}} \rangle  \cdot  \langle {\bm  S_{2}} \rangle }{m_1m_2}  + \dfrac{\vartheta}{m_1m_2}   \Big( \langle {\bm  S_{1}} \rangle \times \langle {\bm  S_{2}} \rangle \Big) \cdot \bm \nabla  \right]  \delta^{(3)}(\bm x)  \nonumber \\
     &\; + \dfrac{\delta_1 \delta_2}{4 \pi |\bm x|} \left[1 + \dfrac{\bm \ell^2}{4m_1m_2} - \bm \ell^2   \left( \dfrac{1}{m_1^2}+ \dfrac{1}{m_2^2}\right)\right]  - \dfrac{\vartheta}{4 \pi |\bm x|}  \dfrac{\bm \ell^2}{m_1m_2} \bm \ell \cdot \Big(\langle {\bm  S_{2}} \rangle \delta_1 - \langle {\bm  S_{1}} \rangle \delta_2  \Big)  \nonumber \\
     &\; + \dfrac{ \bm Q_{ij}}{4 \pi | \bm x|^2} \dfrac{\langle {\bm  S_{1i}} \rangle  \langle {\bm  S_{2j}} \rangle}{m_1m_2} - \dfrac{\bm L }{4 |\bm x|^3} \! \cdot \! \bigg[ \delta_1  \langle {\bm  S_{2}} \rangle \left(\dfrac{3}{4m_2^2} + \dfrac{1}{m_1m_2} \right)   + 1\leftrightarrow 2   \bigg]  +  \dfrac{\vartheta \bm \ell^2}{4 \pi m_1m_2 |\bm x|^3} \bm x \cdot \Big( \langle {\bm  S_{2}}  \rangle  \times  \langle {\bm  S_{1}} \rangle \Big)  \bigg\} ,
\end{align}
\end{widetext}
where $\bm L \! =\! \bm x \!\times \! \frac{\bm \ell}{2}$ is the angular momentum and $\bm Q_{ij} \!= \!\delta_{ij } \!-\! 3 \frac{\bm x_i \bm x_j}{ \bm |\bm x|^2}$.
As anticipated in Section \ref{sec5}, the linear contributions in the $\vartheta$ parameter are explicitly calculated and we pay attention to the Coulomb-like term in $\vartheta$ with dependence on the spin. Also, it is worth to highlight the spin correction coming from the $\vartheta$-contribution to the contact term.
In the case of the $e^-e^-$ scattering, $m_1=m_2=m_e$ and $e_1=e_2=e$, with $m_e$ and $e$ being the electron mass and charge, respectively. Then, in the situation of an  $e^-e^-$ scattering with spin-flip, both $\delta_1 $ and $\delta_2$ are trivial.  In this sense, the potential given by Eq. \eqref{potentialV} assumes the form 
\begin{align} \label{potentialVflip}
    V(\bm x) =
    &\;   \dfrac{e^2}{4\pi |\bm x|^2} \left[ \dfrac{  \bm Q_{ij} \langle {\bm  S_{1i}} \rangle  \langle {\bm  S_{2j}} \rangle \! + \! \vartheta \bm \ell^2 \bm{\hat{x}} \! \cdot \! \Big(\! \langle {\bm  S_{2}}  \rangle \! \times \!  \langle {\bm  S_{1}} \rangle \! \Big)}{ m_e^2}  \right] \nonumber \\
      &\;  + \dfrac{\vartheta e^2 }{m_e^2} \Big( \langle {\bm  S_{1}} \rangle \times \langle {\bm  S_{2}} \rangle \Big) \cdot  \bm \nabla   \delta^{(3)}(\bm x) .
\end{align}
In the case where the  $e^-e^-$ scattering does not have the spin-flip, $\delta_1 =\delta_2 $ and $ \langle {\bm  S_{1}} \rangle  =  \langle {\bm  S_{2}} \rangle$, and the  Eq. \eqref{potentialV} reduces to 
\begin{align} \label{potentialVnoflip}
    V (\bm x) =&\; \dfrac{e^2 }{4\pi |\bm x|} \left(1 - \dfrac{7\bm \ell^2}{4m_e^2}  \right) \delta_1^2 \nonumber \\
     &\; +    \dfrac{e^2}{m_e^2} \bigg[ \left( \dfrac{5 \delta_1 \delta_2 }{8}   -  \dfrac{2\langle {\bm  S_{1}} \rangle  \cdot  \langle {\bm  S_{2}} \rangle }{3} \right)  \delta^{(3)}(\bm x)  \nonumber \\
     &\;   + \dfrac{ \bm Q_{ij} \langle {\bm  S_{1i}} \rangle  \langle {\bm  S_{2j}} \rangle}{4 \pi | \bm x|^2}  -   \dfrac{7 \delta_1 }{8 |\bm x|^3}  \bm L \cdot   \langle {\bm  S_{2}} \rangle    \bigg] \;  .
\end{align}
The contribution from LQG arises specifically in scenarios involving scattering with spin-flip, as outlined in Eq. \eqref{potentialVflip}. Notably, this contribution does not take into account the typical Coulomb-like term. The Coulombian interaction, being a gauge interaction, preserves the chirality of the particles. Hence, it follows that, in the absence of spin-flip, as Eq. \eqref{potentialVnoflip} shows, Coulomb potential comes out. 
\section{The coupling with the EM field and the non-relativistic Hamiltonian} \label{sec7}
The gauge principle applied to the Eq. (\ref{ActionDirac1}) means to introduce the minimal coupling with the EM field via the usual redefinition of the $4$-momentum $p_{\mu} \longmapsto p_{\mu} - e \, A_\mu$ in which the modified Dirac Eq. \eqref{Dirac1} is given by
\begin{align}\label{EqpA}
\left[ \gamma^\mu\left(p_\mu \! - \! e A_\mu \right) +\vartheta \left(\bm{p}-e\bm{A}\right)^2 \gamma^0 \gamma_5 \!- \! m  \right] u(p) =0 \; .
\end{align}
Using the previous notation for the $u$-amplitude in the momentum space, {\it i.e.}, $u= (u_a \; \; \; u_b)^t$, the system of Eqs. in (\ref{EqpA}) is read as
\begin{subequations}
\begin{align}
& \!\!\!
\left( E \! - \! e\phi \! - \! m \right) \! u_a \! - \! \left[\bm{\sigma} \! \cdot \!  \left(\bm{p} \! - \! e \bm{A} \right)  \! - \! \vartheta \left(\bm{p} \! - \! e\bm{A}\right)^2 \right] \! u_b \! = \! 0  ,
\label{s1}
 \\
& \!\!\!
\left( E \! - \! e\phi \! + \! m \right) \! u_b \! - \! \left[\bm{\sigma} \! \cdot \! \left(\bm{p} \! - \! e \bm{A} \right) \! - \! \vartheta\left(\bm{p} \! - \! e\bm{A}\right)^2 \right] \! u_a \! = \! 0 .
\label{s2}
\end{align}
\end{subequations}
From (\ref{s2}), we have the relation
\begin{equation}\label{ub}
u_{b} = \dfrac{\bm{\sigma}\cdot  \left(\bm{p} - e \bm{A} \right) -\vartheta\left(\bm{p}-e\bm{A}\right)^2 }{E-e\phi+m} \, u_a \; ,
\end{equation}
that allows us to write the spinor $u(p)$ in the form
\begin{align}\label{uua}
u(p) =
\begin{pmatrix}
       \mathds{1} \\[0.2cm]
        \dfrac{\bm{\sigma}\cdot \left(\bm{p} - e \bm{A} \right) -\vartheta\left(\bm{p}-e\bm{A}\right)^2 }{E-e\phi +m}
\end{pmatrix}
u_{a} \; .
\end{align}
In the non-relativistic  regime, the linear momentum is $ \bm{p} \ll m$, and the dominant energy term is of the order of the rest energy of the particle, that is, $E \approx  m$. The kinetic and potential energy are much smaller than the rest energy, that is, the approximation $e\phi \ll m$
is valid in the previous relations. Therefore, the relations between the Eqs. (\ref{ub}) and (\ref{uua}) are simplified as
\begin{subequations}
\begin{eqnarray}
u_b  \!&\approx&\!  \dfrac{\bm{\sigma} \cdot \bm{\pi}  -\vartheta\bm{\pi}^2 }{2m} \, u_a \; ,
\label{ubapprox}
\\
u_{\textrm{\tiny NR}}(p) \!&\approx&\!
\begin{pmatrix}
\mathds{1} \\[0.2cm]
\dfrac{\bm{\sigma}  \cdot \bm{\pi} -\vartheta\bm{\pi}^2 }{2m}
\end{pmatrix} u_{a} \; ,
\end{eqnarray}
\end{subequations}
where $\bm{\pi} \equiv \bm{p} - e \bm{A}$. Substituting Eq. \eqref{ubapprox} in Eq. \eqref{s1}, the Eq. \eqref{s1} is so written in the form $E_{\textrm{\tiny NR}}u=(E-m)u=Hu$, in which $ E_{\textrm{\tiny NR}} $ is the energy of a particle, and $H$ is the Hamiltonian operator in the NR limit:
\begin{align}
H = e\phi +  \dfrac{\left(\bm{\sigma}  \cdot \bm{\pi}  -\vartheta\bm{\pi}^2 \right)^2 }{2m} \, .
\end{align}
The kinetic term yields the operator
\begin{equation}
\left(\bm{\sigma} \! \cdot \! \bm{\pi} \! - \! \vartheta\bm{\pi}^2 \right)^2 \! = \! \left(\bm{\sigma} \! \cdot \! \bm{\pi}  \right)^2 \! + \! \vartheta^2 \bm{\pi}^4 \! - \! \vartheta\bm{\pi}^2  \left(\bm{\sigma}  \cdot \bm{\pi}  \right) \! - \! \vartheta \left(\bm{\sigma} \!  \cdot \! \bm{\pi}  \right) \bm{\pi}^2  ,
\end{equation}
in which new terms as $\bm{\pi}^4$ and $\bm{\pi}^2  \left(\bm{\sigma}  \cdot \bm{\pi} \right)$ emerge with the $\vartheta$-parameter. We use the following relations
\begin{subequations}
\begin{eqnarray}
\left(\bm{\pi} \cdot \bm{\sigma} \right)^2 \!&=&\! \left(\bm{p} - e\bm{A}\right)^2 -e \bm{\sigma} \cdot \bm{B} \; ,
\\
\bm{\pi}^2  \left( \bm{\sigma} \cdot \bm{\pi}  \right) \!&=&\! \left(\bm{\sigma} \cdot \bm{\pi}  \right) \bm{\pi}^2   +  e \bm{\sigma} \cdot \partial_t \bm{E}
\nonumber \\
&&
 +\, 2 e^2 \bm{\sigma}  \cdot  \left[  \bm{A} \times \left(\bm{p} \times  \bm{A} \right) \right] \; ,
\end{eqnarray}
\end{subequations}
for a magnetic field $({\bf B})$, in which $\bm{B} = \bm \nabla \times \bm{A}$. Therefore, the Hamiltonian operator is
\begin{align}
H =&\;  e\phi +\dfrac{\left(\bm{p} - e\bm{A}\right)^2 }{2m} -\dfrac{e }{2m}  \bm{\sigma} \cdot \bm{B} \nonumber \\
&\; -\dfrac{\vartheta}{m} \left(\bm{\sigma} \cdot \bm{\pi}  \right)  \left(\bm{p} - e\bm{A}\right)^2
-\dfrac{\vartheta \, e }{2m}   \bm{\sigma} \cdot  \partial_t \bm{E}
\nonumber \\
&\; -\dfrac{\vartheta e^2}{m} \, \bm{\sigma} \cdot \left[  \bm{A} \times \left(\bm{p} \times  \bm{A} \right) \right] +\dfrac{\vartheta^2 }{2m}  \bm{\pi}^4 \; .
\end{align}
In the particular case of a weak and homogeneous magnetic field, the vector potential is
$\bm{A} = (\bm{B} \times \bm{x})/2$, and we can neglect quadratic terms in $\bm{A}$. Thereby,
we can approximate the terms
\begin{subequations}
\begin{eqnarray}
&&
\bm{\pi}^2
\approx \bm{p}^2 -e (\bm{B} \cdot  \bm{L}) \; ,
\\
&&
\bm{\pi}^4
\approx \bm{p}^4 - 2e \bm{p}^2 \left(\bm{B} \cdot \bm{L} \right) \; ,
\\
&&
\bm{\sigma} \cdot \left[  \bm{A} \times \left(\bm{p} \times  \bm{A} \right) \right] \approx 0 \; ,
\\
&&
\bm{\sigma}\cdot (\bm{p} -e \bm{A})(\bm{p} -e \bm{A})^2 \approx  - e  \bm{p}^2 \bm{B} \cdot (\bm{r} \times \bm{S})
\nonumber \\
&&
+\, \frac{1}{2}(\bm{S}\cdot \bm{p}) \left( \bm{p}^2 -e \bm{B} \cdot  \bm{L} \right)\; ,
\end{eqnarray}
\end{subequations}
where we have used that $\bm{S}=\bm{\sigma}/2$. Under these conditions, the Hamiltonian operator is
\begin{align}\label{HNRB}
H =&\; e \phi + \dfrac{\bm{p}^2}{2m} \Big[1  + 2\vartheta  \big( e\bm{B}\cdot (\bm{x}\times \bm{S}) - 2  \bm{S}\cdot \bm{p} \big) 
 \nonumber \\
&\; + \vartheta^2 \left( \bm{p}^2  - 2 \bm{B}\cdot \bm{L} \right)  \Big] -\dfrac{e }{2m}  \left(\bm{L}+2 \bm{S} \right)\cdot \bm{B}  \nonumber \\
&\; -\dfrac{\vartheta \, e }{m} \, \bm{S} \cdot  \partial_t \bm{E}
+ \dfrac{2\vartheta\,e}{m} \left(\bm{B} \cdot \bm{L}\right) \left( \bm{S}\cdot \bm{p} \right) .
\end{align}
The term $\bm{S} \cdot \partial_t \bm{E}$ is projection of the electric field on the spin direction. It emerges from the Faraday law, and expresses the electric dipole momentum variable with the time. In the case of a electrostatic field, as the Coulomb electric field, this term is null. The other terms have contribution of the uniform magnetic field, of the linear momentum, and also of angular momentum $({\bf L})$ for the NR particle.
\section{Renormalizability} \label{sec8}
The presence of a $\bm\nabla^2$ in the modified Dirac action introduces a higher-order space derivative.  Besides gauge symmetry and causality, which are fundamental ingredients of a consistent quantum field theory, renormalizability can be also checked in this model. From the discussions in Section \ref{sec2} and \ref{sec4}, respectively, we see that the familiar causal structure of special relativity may not be such a rigid rule  that must hold in LSV scenarios, and that $U(1)$ gauge symmetry is ensured if minimal coupling with the EM field is introduced. The missing point, regarding renormalizability, is a relevant discussion and let us take into consideration the fact that the modified fermion propagator given by Eq. \eqref{propagador} becomes more convergent with respect to the usual case of QED. Thus, we conclude that the model is super-renormalizable. This is the matter discussed in the present Section.
 To start our inspection on the power-counting of the primitively divergent diagrams , we  remember that the ultraviolet behavior of the photon propagator and the modified fermion propagator are proportional to $\sim \tfrac{1}{p^2}$. It is important to pay attention to this modified ultraviolet behavior of the fermion propagator. For the power counting, we denote the internal (external) photon and fermion lines by $I_A (E_A)$ and $I_\psi (E_\psi)$, respectively.
We also denote the three-vertices of the model, Eq.  \eqref{QEDaction}, by $V_{\psi\psi A}$, $V_{\psi\psi A \partial}$, and $V_{2\psi 2A}$. $V_{\psi\psi A \partial}$ stands for the derivative fermion-fermion-photon vertex. In what follows, $V \equiv V_{\psi\psi A} +V_{\psi\psi A \partial}+ V_{2\psi 2A} $ gives the total number of vertices of the considered diagram.
           Thus, the superficial degree of divergence can be read off from the expression
           \begin{align} \label{superficial}
               \delta_{\tiny \textrm{graph}} = 4(I_A +  I_\psi)  - 4 (V -1) -2(I_A+I_\psi) + V_{\psi\psi A \partial} ,
           \end{align}
           where  the first term comes from the integration elements of all internal lines, the second term is related with the Dirac delta functions associated to the vertices, but taking into account that one of them remains to express the overall energy-momentum  conservation. The third term is due to the propagators (and here the fermion propagator is modified) and the last term corresponds to the momentum appearing in the derivative  fermion-fermion-photon vertex. 

        We also have to consider the following topological relations
           \begin{align}
               2 I_A + E_A
               &=  V_{\psi\psi A}+ V_{\psi\psi A \partial} + 2V_{2\psi 2A} , \\
               2 I_\psi + E_\psi &=  2V_{\psi\psi A}+ 2 V_{\psi\psi A \partial} + 2V_{2\psi 2A} .
           \end{align} 
         By substituting these topological relations into the expression for the superficial degree of divergence, Eq. \eqref{superficial}, we obtain
           \begin{align}
               \delta_{\tiny \textrm{graph}} = 4 - E_A - E_\psi -V_{\psi\psi A}. 
           \end{align}
where  $V_{\psi\psi A}$ is the usual QED vertex with two fermionic lines, one photonic line, and no derivative present. 
Since that $V_{\psi\psi A}$ appears with a negative coefficient, the conclusion is in favor of a super-renormalizable model. This result is a consequence of the more convergent behavior of the modified fermion propagator in the ultraviolet regime, as given by Eq. \eqref{propagador}. Though the LQG-corrected vertices carry the $\vartheta$-parameter with negative mass dimension, the true coupling constant of the model is the (dimensionless) fundamental charge, $e$. This is why the power-counting rule above does not indicate non-renormalizability.
Before closing this Section, we would like to call into question a subtle point in connection with the discussion of renormalizability. The presence of the $\vartheta$-term with a $\bm p^2$-dependence in the fermionic free action is responsible for a better convergent ultraviolet behavior of the fermion field propagator (given by Eqs. \eqref{propagador} and \eqref{RDf}), which falls off as $\tfrac{1}{\bm p^2}$, instead of $\tfrac{1}{| \bm p|}$, yielding then to a final degree of divergence which points to super-renormalizability. However, it is interesting to notice that, if we adopt a different viewpoint, by dropping the $\vartheta$-term from the free fermionic action and rather facing it as an insertion vertex with a $\bm p^2$-dependence, we restore the usual QED fermion propagator back, but the power-counting fails to indicate both renormalizability and super-renormalizability, since we will have a propagator behaving as usually,  $\tfrac{1}{| \bm p|}$ in the ultraviolet regime, and vertex insertions that now introduce $\bm p^2$-factors.
  It seems that we are before conflicting results. Nevertheless, we argue that, in our way of approaching the extended Dirac equation \eqref{eqD1} and the field-theoretic action \eqref{QEDaction}, we are keeping $\vartheta$-corrections to all orders in $\vartheta$ (we refer again to Eqs. \eqref{propagador} and \eqref{RDf}). Actually, the fermion propagator read off from these equations account for the $\vartheta\bm p^2$-vertex inserted to all orders in the usual fermion propagator. On the other hand, the plane wave solutions built up from Eqs. \eqref{solmom}-\eqref{waveeq}, and used throughout our developments in the previous Sections, also keep the $\vartheta\bm p^2$-term as part of the free fermionic wave equation. Following this path is tantamount to saying that, in our treatment with the $\vartheta$-term taken as part of the free Lagrangian, the field-theoretic action \eqref{QEDaction} has the property of being super-renormalizable, as our power-counting rule derived above shows. However, since our investigation here is bound to quantum-mechanical and tree-level field-theoretic considerations, radiative corrections have not been computed and the renormalization procedure lies beyond the scope of the present paper. But, to our sense, it is fair to raise this subtle question and share it with the readers.
\section{Concluding remarks}
\label{sec9}
In this paper, we study the solutions of a modified Dirac equation by the effects of the LQG, and it coupling with the EM field through the $U(1)$ gauge symmetry. The modification of the Dirac action is due to presence of a length scale $(\vartheta)$ introduced by the LQG, where the Dirac kinetic term contains the higher order derivative term $\vartheta\,\bar{\psi}\gamma^{0}\gamma_{5}\,\bm \nabla^2\psi$. We obtain the plane solutions in the momentum space associated with the positive and negative energies for the modified Dirac operator. These solutions express an asymmetry in these solutions due to Lorentz symmetry breaking induced by the $\vartheta$-length scale. This is the consequence of the asymmetry for the frequencies associated with the fermion dispersion relations. The group velocity for fermions is showed for the positive energy solution. The fermion propagator in this LSV scenario is obtained in the momentum space. In the ultraviolet regime, the propagator has an improved behavior in which it goes to zero with the squared inverse of the momentum $(\sim \tfrac{1}{p^2})$. This behavior of the fermionic propagator is responsible for a conclusion in favor of a super-renormalizable model by power counting.
The $U(1)$ global symmetry in the LQG Dirac action motivates us to calculate the correspondent conserved current through the Noether theorem.
The term $\vartheta\,\bar{\psi}\gamma^{0}\gamma_{5}\,\bm \nabla^2\psi$ with the Laplacian operator implies that the spatial-component for the
conserved current is altered with new terms that depend on the ${ \vartheta}$-parameter, whereas that time-component is unaltered.
The next step is so to investigate the $U(1)$ local symmetry of this theory. The minimal coupling with the EM field preserves the $U(1)$ local transformation, that leads us to a QED emerged from the LQG. Thereby, we study the coupling of the conserved current with the EM field through the Gordon identity. New terms emerge in the current that can be interpreted as contributions to the lepton's electric dipole momentum due exclusively to the ${ \vartheta}$-length scale. Afterwards, we obtain the non-relativistic limit in the Dirac equation, where the Hamiltonian for the NR particle acquires terms that depend on the spin and orbital angular momentum projected on the uniform magnetic field, and also the projection on the particle linear momentum. A new term that depends on the interaction of the electric dipole momentum  emerges in the NR limit and an explicit calculation of the EDM shows that it is within the experimental limit and very close to the current Standard Model predictions. All these new effects are removed whenever $\vartheta \rightarrow 0$.

Also, to render clearer the terms that appear after carrying out the Gordon decomposition of the electric magnetic current, we have devoted special attention to the calculation of the $e^{-}e^-$ interparticle potential and the linear corrections in the $\vartheta$ parameter shows us how LQG effects correct the usual spin- and velocity-dependent contributions present in the potential.      

To end up, we briefly investigate the power-counting renormalizability of the model. Since the fermion propagator exhibits a more convergent ultrared behavior, $D^{-1} \sim \tfrac{1}{p^2}$, the model becomes super-renormalizable. Thus, this conclusion motivates us to investigate new effects of this modified QED, such as the corrections to the $g-2$ factor for the electron and muon, which are decisive precision measurements in QED. Furthermore, the study of the scalar sector and of the EM modified by the LQG, as described in the Ref. \cite{Li23}, are interesting challenges for a forthcoming project. As a perspective, it is worth noting that one can explore the effects of the Hamiltonian in Eq. \eqref{HNRB} on spectral lines, e.g., of the Hydrogen atom. 
\section{Acknowledgements}
J.P.S.M. and J.M.A.P. express their gratitude to CNPq-Brazil for granting their PhD Fellowships.   The authors are grateful to the referee  for helping to clarify some points of the work.
\end{document}